\documentclass[reprint,superscriptaddress,amsmath,amssymb,aps,prb,longbibliography,
]{revtex4-2}

\usepackage[dvipsnames]{xcolor}
\definecolor{mblue}{RGB}{31, 119, 180}
\usepackage{hyperref}
\hypersetup{backref,%pdfpagemode=FullScreen,
colorlinks=true,breaklinks,urlcolor=mblue,linkcolor=mblue,citecolor=mblue}
\usepackage{physics, braket, bm, amsthm, amsmath, amssymb, mathrsfs, graphicx, dcolumn, subfigure}
\usepackage{mathrsfs}
\usepackage[]{ulem}

\begin{document}

\title{Machine-learning-inspired quantum optimal control\\ of nonadiabatic geometric quantum computation via reverse engineering}
%\title{Neural-network-based optimal quantum control \\ of nonadiabatic geometric quantum computation via reverse engineering}
\author{Meng-Yun Mao}
\affiliation{College of Physics, Nanjing University of Aeronautics and Astronautics, Nanjing 211106, China}
\affiliation{Key Laboratory of Aerospace Information Materials and Physics (NUAA), MIIT, Nanjing 211106, China}

\author{Zheng Cheng}
\affiliation{College of Physics, Nanjing University of Aeronautics and Astronautics, Nanjing 211106, China}
\affiliation{Key Laboratory of Aerospace Information Materials and Physics (NUAA), MIIT, Nanjing 211106, China}

\author{Yan Xia}
\email{xia-208@163.com}
\affiliation{Fujian Key Laboratory of Quantum Information and Quantum Optics, Fuzhou University, Fuzhou 350116, China}
\affiliation{Department of Physics, Fuzhou University, Fuzhou 350116, China}

\author{Andrzej M. Ole\'s}
 \email{a.m.oles@fkf.mpi.de}
\affiliation{\mbox{Max Planck Institute for Solid State Research,
             Heisenbergstrasse 1, D-70569 Stuttgart, Germany} }
\affiliation{\mbox{Institute of Theoretical Physics, Jagiellonian University,
             Prof. Stanis\l{}awa \L{}ojasiewicza 11, PL-30348 Krak\'ow, Poland}}

\author{Wen-Long You$\,$}
\email{wlyou@nuaa.edu.cn}
\affiliation{College of Physics, Nanjing University of Aeronautics and Astronautics, Nanjing 211106, China}
\affiliation{Key Laboratory of Aerospace Information Materials and Physics (NUAA), MIIT, Nanjing 211106, China}

\date{\today}
%%%%%%%%%%%%%%%%%%%%%%%%%%%%%%%%
%%%%%%%%%%% ABSTRACT %%%%%%%%%%%
%%%%%%%%%%%%%%%%%%%%%%%%%%%%%%%%
\begin{abstract}
Quantum control plays an irreplaceable role in practical use of quantum computers.
However, some challenges have to be overcome to find more suitable and diverse control parameters.
We propose a promising and generalizable average-fidelity-based machine-learning-inspired method to optimize the control parameters,
in which a neural network with periodic feature enhancement is used as an ansatz.
In the implementation of a single-qubit gate by cat-state nonadiabatic geometric quantum computation via reverse engineering,
compared with the control parameters in the simple form of a trigonometric function,
our approach can yield significantly higher-fidelity ($>99.99\%$) phase gates,
such as the $\pi / 8$ gate (T gate).
Single-qubit gates are robust against systematic noise,
additive white Gaussian noise and decoherence.
We numerically demonstrate that the neural network possesses the ability to expand the model space.
With the help of our optimization,
we provide a feasible way to implement cascaded multi-qubit gates with high quality in a bosonic system.
Therefore,
the machine-learning-inspired method may be feasible in quantum optimal control of nonadiabatic geometric quantum computation.
\end{abstract}

% \pacs{}
%\keywords{}

\maketitle


%%%%%%%%%%%%%%%%%%%%%%%%%%%%%%%%
%%%%%%%%% INTRODUCTION %%%%%%%%%
%%%%%%%%%%%%%%%%%%%%%%%%%%%%%%%%
\section{Introduction}

Multi-qubit gates are widely used in quantum circuits~\cite{10.5555/3179553.3179560, PhysRevLett.93.130502, liu_universal_2022},
quantum error correction~\cite{reed_realization_2012, taminiau_universal_2014, PRXQuantum.2.030345}, and other fields
\cite{PhysRevA.32.3266, feynman_quantum_1985, chatterjee_transmon-based_2015, fan_efficient_2021}.
Single-shot multi-qubit gates~\cite{baker2022single, lanyon_simplifying_2009, PhysRevLett.102.040501}
%refer to
specify quantum circuits that evolute in a well-controlled,
uninterrupted, and continuous-time way %to realize the
%quantum gate
to implement the quantum computation~\cite{spiteri_quantum_2018}.
Compared to the cascaded gates,
single-shot multi-qubit gates can greatly reduce the circuit depth and shorten the implementation time,
thus suppressing decoherence~\cite{PhysRevA.102.012601, kim_high-fidelity_2022}.
However, the single-shot method is difficult to realize in experiments
owing to the restricted conditions of simultaneously manipulating
multiple physical systems and building complex couplings.

One of the ways to mitigate the above difficulty is the application of
single- and two-qubit gates to equivalently implement the function of
multi-qubit gates~\cite{he_decompositions_2017, PhysRevA.106.042602, PRXQuantum.2.040348},
and the decomposition is guaranteed by the Solovay-Kitaev theorem~\cite{nielsen2002quantum}.
Although the decomposition method loses the upper hand in terms
of the circuit depth and the implementation time, it has a wider scope
of application due to direct execution on the quantum processing unit,
namely, the brain of a quantum computer~\cite{nakanishi2021quantum}.
The realization of the synthetic gates depends on the universal single-qubit gates and a two-qubit entangling gate with high fidelity~\cite{kim_high-fidelity_2022}. However,
statistical imprecision in the experimental controls,
interactions between the system and the environment,
and random driving forces from the environment will cause a reduction in fidelity~\cite{lidar2013quantum}.

Universal single-qubit gates based on the geometric phase in quantum systems have recently shown robustness against control-parameter fluctuations~\cite{cohen_geometric_2019}.
An adiabatically evolving system driven by a nondegenerate Hamiltonian exhibits geometric phase under cyclic evolution~\cite{anandan_geometric_1992, berry1988geometric, berry1984quantal}.
The geometric phase arising from cyclic evolution of quantum systems is uniquely determined by the geometric structure of the enclosed path in the parameter space~\cite{ZHANG20231},
which is the well-known analog of the rotational effect in differential geometry when a vector is parallel transported~\cite{bohm2003geometric, shapere1989geometric}.
%An adiabatically evolving system driven by a nondegenerate Hamiltonian exhibits geometric phase under cyclic evolution~\cite{anandan_geometric_1992, berry1988geometric, berry1984quantal}.
However, %the strict limit of the adiabatic condition
the strict condition of the adiabatic limit requires the evolution time to be infinitely long,
which inevitably gives rise to decoherence of the system~\cite{PhysRevLett.87.097901}.
The nonadiabatic geometric phase gets rid of the bondage of the adiabatic condition,
making it possible to shorten the evolution time of the system to a great extent.
The nonadiabatic geometric phase lays a solid foundation for nonadiabatic geometric quantum computation (NGQC).
Recently, NGQC has been executed theoretically~\cite{PhysRevA.67.052309, Sjoqvist_2012, PhysRevA.96.052316, PhysRevApplied.10.054051} and experimentally~\cite{PhysRevLett.110.190501, PhysRevLett.124.230503, PhysRevLett.89.097902, PhysRevLett.122.010503, PhysRevLett.127.030502}
 in multiple quantum systems.
Later, NGQC has been further promoted to NGQC+~\cite{liu_plug-and-play_2019}.
The NGQC+ scheme loosens the conditions for the realization of NGQC to a certain extent, which becomes more compatible with optimal control methods.
%Especially the extension makes NGQC more compatible with optimal control methods, such as
Several schemes have been developed, including
counter-adiabatic driving~\cite{PhysRevLett.105.123003, zhang_fast_2015},
dynamical decoupling~\cite{PhysRevLett.95.180501, PhysRevLett.106.240501},
and machine-learning-based optimization techniques~\cite{PhysRevA.97.042324, PhysRevA.97.052333}.

Recent researches have shown that logical qubit encoding is promising to protect quantum computation from errors
~\cite{harrington_engineered_2022, campbell_roads_2017, blais_quantum_2020}.
However, in standard logical qubit systems based on multiple physical qubits~\cite{PhysRevLett.81.2152, schindler2011experimental, chiaverini_realization_2004, PhysRevA.101.012316},
quantum error correction and logical operations are difficult to achieve because the number of error channels rapidly increases with the number of qubits~\cite{hu_quantum_2019}.
For the realization of logical qubits,
bosonic systems are promising candidates,
because the number of error channels can dramatically drop~\cite{hu_quantum_2019, PhysRevA.106.022431}
with taking advantage of the infinite-dimensional Hilbert space of the harmonic oscillator.
The cat states of bosons have been widely used in quantum computation and quantum error correction~\cite{ma_error-transparent_2020, gertler_protecting_2021, Albert_2019, cai_bosonic_2021}.
Encoding in cat-state subspace via reverse engineering provides a feasible scheme for the realization of NGQC in a bosonic system~\cite{kang_nonadiabatic_2022}.
The application of reverse engineering,
which constructs the Hamiltonian based on the corresponding invariant,
makes it easier to find more free parameters to control the evolution path~\cite{kang_flexible_2020, PhysRevA.89.043408}.
Numerous studies have demonstrated that the %evolution
tentative form of control parameters shapes the time evolution of quantum systems in a potentially useful way~\cite{PhysRevLett.112.240503, rice2000optical, shapiro2003principles}.
Typical forms of control parameters include polynomials of trigonometric functions~\cite{li_invariant-based_2021, Wang_2023},
as well as the product form of the trigonometric and complex exponential functions~\cite{PhysRevA.102.022617, liu_optimized_2021}.
In the system to be elucidated subsequently, the control parameters are limited to simple trigonometric functions.
%In the system to be discussed below,
%the control parameters are merely simple trigonometric functions.
The %selection
adjustment
of the evolution form of control parameters is of great importance in quantum computation.

Adopting the machine-learning technology and optimization theory has been proved to be applicable to
%finding the ground state and
optimizing the control parameters of variational states
in a variety of interacting quantum many-body systems
~\cite{carleo_solving_2017, gao_efficient_2017, hartmann_neural-network_2019, schmitt_quantum_2020, dborin_simulating_2022}. %~\cite{machnes_comparing_2011, spiteri_quantum_2018, khaneja_optimal_2005}.
Although designing control parameters to acquire high-fidelity quantum gates by neural network has been extensively studied for a long time~\cite{machnes_comparing_2011, spiteri_quantum_2018},
it is still a flourishing and attractive research topic.
Researchers designed dispersed and aperiodic control parameters by gradient ascent pulse engineering (GRAPE) under the guidance of state fidelity in nuclear magnetic resonance~\cite{khaneja_optimal_2005}.
Here we introduce this method into the bosonic system, where
the aperiodic discontinuous function is generalized to the periodic continuous function. %\red{, and we combine GAPE with the neural network}.
We find that the incorporation of GRAPE enables the neural network to possess a powerful representation ability,
which can expand the model space through the nonlinear activation function to fit any smooth periodic function and aperiodic function~\cite{dong_method_2021, ziyin_neural_2020}.
As a result,
we optimize continuous and periodic control parameters,
which are easier to physically implement,
through the neural network with the enhancement of periodic characteristics.

The rest of the paper is organized as follows. In
Sec.~\ref{NGQC} we revisit the NGQC+ with cat states via reverse engineering.
Section~\ref{ANN} is devoted to the construction of the neural network guided by the average fidelity with periodic feature enhancement to improve the performance of single-qubit gates.
In Sec.~\ref{results},
we benchmark the optimization on the T gate ($\pi / 8$ gate) and demonstrate that the neural network can effectively expand the model space.
Furthermore, we assess the performance of the protocol under systematic noise, random noise and decoherence effect via numerical simulations.
Finally, the conclusions and outlook are given in Sec.~\ref{conclusion}.

%\newpage

\section{NGQC+ with cat states based on reverse engineering} \label{NGQC}
%\subsection{NGQC+}
Applying reverse engineering to quantum computation not only permits the Hamiltonian to be more physically realizable,
but also makes the implementation of quantum gates more flexible~\cite{kang_flexible_2020, puri_engineering_2017, kang_pulse_2018}.
Consider a time-dependent Hamiltonian $H(t)$ and the corresponding dynamic invariant $I(t)$,
which satisfies the following equation~\cite{lewis_exact_1969} ($\hbar = 1$):
\begin{equation}
    \label{eq:H_I}
    i \frac{\partial}{\partial t} I(t) - [H(t) , I(t)] = 0.
\end{equation}
To realize NGQC+,
we select a set of time-dependent eigenstates $| \phi_{l}(t) \rangle$ $(l =1, 2,\cdots, d)$ of $I(t)$ to span a $d$-dimensional computational subspace $\mathcal{S}$,
which are supposed to satisfy the three conditions below~\cite{liu_plug-and-play_2019}.
First,
the computational basis should satisfy the boundary conditions at times $t = 0$ and $L$, i.e.,
$| \phi_{l}(0) \rangle = | \phi_{l}(L) \rangle$, to ensure that the evolution is cyclic.
Here, $L$ is the evolution period.
Secondly,
we can rewrite Eq.~\eqref{eq:H_I} based on eigenvectors of %$H(t)$
$I(t)$ as
\begin{equation}
    \dot{\Xi}_{l}(t) = - i [H(t) , \Xi_{l}(t)],
\end{equation}
where $\Xi_{l}(t) = | \phi_{l}(t) \rangle \langle \phi_{l}(t) |$ is the projective operator of $| \phi_{l}(t) \rangle$.
Finally, the cumulative dynamic phase of one cycle needs to vanish,
\begin{equation}
    \label{eq:dp}
    \Phi_{l} (L) = - \int_{0}^{L} dt \langle \phi_{l}(t) | H(t) | \phi_{l}(t) \rangle = 0 .
\end{equation}
This condition is the relaxation of parallel transportation $\langle \phi_{l}(t) | H(t) | \phi_{k}(t) \rangle = 0$ in NGQC.

When the conditions of NGQC+ are all satisfied,
the time evolution operator at the final time $t = L$ in subspace $\mathcal{S}$ can be described as
\begin{equation}
    \label{eq:te}
    U(L , 0) = \sum_{l} \exp[i \Theta_{l} (L)] \Xi_{l}(0),
\end{equation}
where $\Theta_{l}(L)$ is the geometric phase,
given by
\begin{equation}
    \label{eq:gp}
    \Theta_{l} (L) = \int_{0}^{L} dt \langle \phi_{l}(t) | i \frac{\partial}{\partial t} | \phi_{l}(t) \rangle .
\end{equation}

%\subsection{Lie algebra}
Suppose a Hamiltonian can be represented as follows,
\begin{equation}
    H(t) = \sum_{j = 1}^{g} \lambda_{j}(t) G_{j},
\end{equation}
where $g$ is the rank of the group and $\{G_{j}\}$ is a group of Hermitian generators of Lie algebra~\cite{PhysRevA.89.043408, kaushal_dynamical_1981, kang_flexible_2020},
%having
obeying the following relations:
\begin{equation}
    [G_{i} , G_{j}] = i \sum_{k} \mu_{ij}^{k} G_{k}, \quad (i , j , k \in \{ 1 , 2 , \cdots , g\}),
\end{equation}
where $\mu_{ij}^{k}$  is the corresponding structure constant.
If an invariant can be written as
\begin{equation}
    I(t) = \sum_{j = 1}^{g} \xi_{j}(t) G_{j}.
\end{equation}
According to Eq.~\eqref{eq:H_I},
it yields
\begin{equation}
    \label{eq:re}
    \dot{\xi}_{k} (t) = \sum_{i , j = 1}^{g} \lambda_{i}(t) \xi_{j}(t) \mu_{ij}^{k}.
\end{equation}
 %If we know
Once $\{ \xi_{j}(t) \}$ are known,
we can thus obtain $\{ \lambda_{j} (t)\}$ according to Eq.~\eqref{eq:re}.

%\subsection{NGQC+ with cat states}
We consider a system in which a resonant single-mode two-photon drive is applied to a Kerr nonlinear resonator.
In the rotating frame,
the system Hamiltonian~\cite{puri_engineering_2017, grimm_stabilization_2020} can be written by
\begin{equation}
    H_{\rm cat} = - K a^{\dagger 2} a^{2} + \epsilon_{2} (e^{2 i \xi} a^{\dagger 2} + e^{- 2 i \xi} a^{2}) ,
\end{equation}
where $K$ is the Kerr nonlinearity,
$a^{\dagger} \ (a)$ is the creation (annihilation) operator of the cavity mode,
$\epsilon_{2}$ is the strength of the two-photon driving,
and $\xi$ is the phase of the driving.
The coherent states $| \pm \alpha \rangle$ with $\alpha = \sqrt{\epsilon_{2} / K} \exp(i \xi)$ are %also
the degenerate eigenstates of $H_{\rm cat}$,
whose superpositions
\begin{equation}
    | \mathcal{C}_{\pm} \rangle = \frac{1}{\sqrt{\mathcal{N}_{\pm}}} (| \alpha \rangle \pm | \textnormal{-} \alpha \rangle),
\end{equation}
are referred to as even (odd) cat states with the normalization constants $\mathcal{N}_{\pm} = 2 \pm 2 \exp(-2 | \alpha |^{2})$.
We apply an external single-photon drive \cite{grimm_stabilization_2020}:
\begin{equation}
  \label{eq:Hct}
    H_{c}(t) = \chi(t) a^{\dagger} a + \epsilon(t) a^{\dagger} + \epsilon^{\ast} (t) a,
\end{equation}
where $\chi(t)$ and $\epsilon(t)$ are the detuning and strength of the driving, respectively.
The total Hamiltonian is described by $H_{tot}(t) = H_{\rm cat} + H_{c}(t)$.
If the constraint that
the energy gaps between cat states and other eigenstates are much larger than $\chi(t)$ and $\epsilon(t)$
is satisfied,
the Hamiltonian can be reduced to two-dimensional subspace spanned by cat states $| \mathcal{C}_{\pm} \rangle$.
The Pauli matrices defined by cat states can be chosen as the Hermitian generators of the Lie group.
The driving Hamiltonian thus can be simplified as $H_{c} = \vec{\Omega}(t) \cdot \vec{\sigma}$,
where $\vec{\Omega}(t) = [\Omega_{x}(t) , \Omega_{y}(t) , \Omega_{z}(t)]$ is a three-dimensional unit vector,
and $\vec{\sigma} = [\sigma_{x} , \sigma_{y} , \sigma_{z}]$.

Consider a dynamic invariant $I(t) = \vec{\zeta}(t) \cdot \vec{\sigma}$,
where $\vec{\zeta}(t) = [\zeta_{x}(t) , \zeta_{y}(t) , \zeta_{z}(t)]$.
Based on Eq.~\eqref{eq:re},
we can get that $\dot{\vec{\zeta}}(t) = 2 \vec{\Omega}(t) \times \vec{\zeta}(t)$ and $| \zeta(t) |$ is constant.
For convenience,
we can let $\vec{\zeta}(t) = (\sin \eta \sin \mu , \cos \eta \sin \mu , \cos \mu)$,
where $\mu$ and $\eta$ are time-dependent control parameters.
The eigenstates of $I(t)$ in the cat-state representation are
\begin{eqnarray}
    | \phi_{+}(t) \rangle & = & \cos \frac{\mu}{2} | \mathcal{C}_{+} \rangle + i \exp(- i \eta) \sin \frac{\mu}{2} | \mathcal{C}_{-} \rangle , \nonumber\\
    | \phi_{-}(t) \rangle & = & i \exp(i \eta) \sin \frac{\mu}{2} | \mathcal{C}_{+} \rangle + \cos \frac{\mu}{2} | \mathcal{C}_{-} \rangle.
\end{eqnarray}
According to Eqs.~\eqref{eq:dp} -~\eqref{eq:gp},
we can calculate the geometric phases
\begin{equation}
    \Theta_{\pm}(L) = \pm \int_{0}^{L} dt \dot{\eta} \sin^{2} \frac{\mu}{2},
\end{equation}
and the dynamic phases
\begin{equation}
    \Phi_{\pm}(L) = \mp \int_{0}^{L} dt \bigg( \frac{1}{2} \dot{\eta} \sin^{2} \mu + \Omega_{z} \bigg) \sec \mu.
\end{equation}
In order to satisfy the conditions
$\Phi_{\pm}(L) = 0$ and $\dot{\vec{\zeta}}(t) = 2 \vec{\Omega}(t) \times \vec{\zeta}(t)$,
we design $\vec{\Omega}(t)$ as
\begin{eqnarray}
  \label{eq:Omega}
    \Omega_{x}(t) & = & \frac{1}{4}[\dot{\eta} \sin \eta \sin (2 \mu) - 2 \dot{\mu} \cos \eta] , \nonumber \\
    \Omega_{y}(t) & = & \frac{1}{4}[\dot{\eta} \cos \eta \sin (2 \mu) + 2 \dot{\mu} \sin \eta] , \nonumber \\
    \Omega_{z}(t) & = & - \frac{1}{2} \dot{\eta} \sin^{2} \mu.
\end{eqnarray}
Therefore,
we set the parameters $\chi(t)$ and $\epsilon(t)$ as
\begin{eqnarray}
   \label{eq:chi-epsilon}
    \chi(t) & = & \frac{\dot{\eta} \sin^{2} \mu \mathcal{N}_{+} \mathcal{N}_{-}}{| \alpha |^{2} (\mathcal{N}_{+}^{2} - \mathcal{N}_{-}^{2})} , \nonumber \\
    \rm{Re} [\epsilon(t)] & = & \frac{\sqrt{\mathcal{N}_{+} \mathcal{N}_{-}}}{4 | \alpha |} (\Omega_{x} \cos \xi - e^{2 | \alpha |^{2}} \Omega_{y} \sin \xi) , \nonumber \\
    \rm{Im} [\epsilon(t)] & = & \frac{\sqrt{\mathcal{N}_{+} \mathcal{N}_{-}}}{4 | \alpha |} (\Omega_{x} \sin \xi + e^{2 | \alpha |^{2}} \Omega_{y} \cos \xi) ,
\end{eqnarray}
which are scarcely different from the forms presented in Ref.~\cite{kang_nonadiabatic_2022}.
Based on Eq.~\eqref{eq:te},
the time evolution operator can be represented as
\begin{eqnarray}
    U(L , 0) =  \left[
        \begin{array}{cc}
            \cos \theta + i \cos \mu_{0} \sin \theta & \exp (i \eta_{0}) \sin \mu_{0} \sin \theta \\
        - \exp(- i \eta_{0}) \sin \mu_{0} \sin \theta & \cos \theta - i \cos \mu_{0} \sin \theta
        \end{array}
        \right], \nonumber \\
        \label{eq:U}
\end{eqnarray}
where $\mu_{0}$ and $\eta_{0}$ are the initial values of $\mu$ and $\eta$, respectively:
\begin{equation}
    \theta = \int_{0}^{L} dt \dot{\eta} \sin^{2} \frac{\mu}{2}.
    \label{eq:theta}
\end{equation}
If we choose different $\mu$, $\eta$, and $\theta$,
we can implement an arbitrary unitary single-qubit gate
 \cite{kang_nonadiabatic_2022}.

\begin{figure}
    \includegraphics[width=\columnwidth]{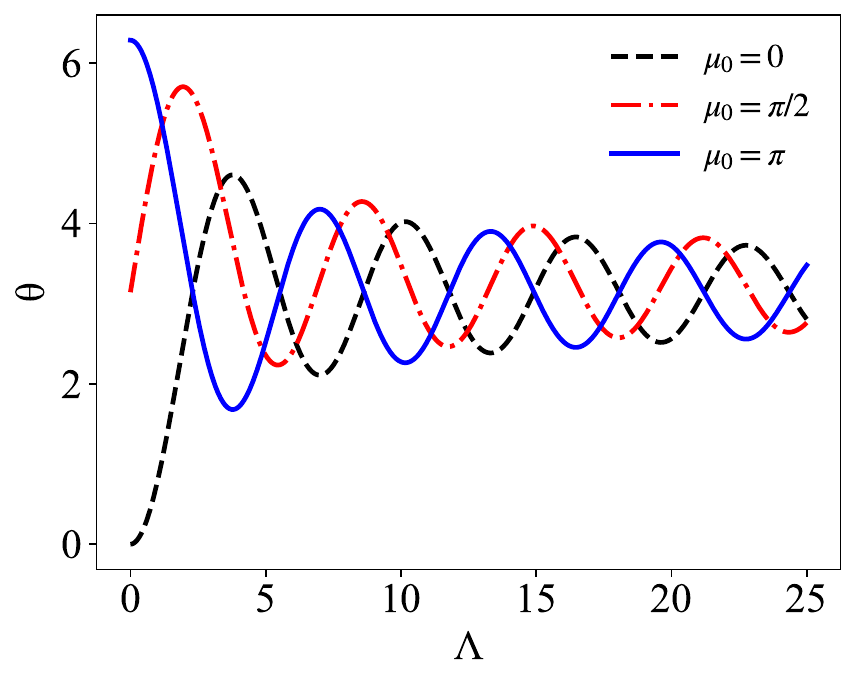}
    \caption{The variation of $\theta$ with respect to $\Lambda$ for $\mu_{0}$=0, $\pi/2$, and $\pi$.
    %$\theta$ is only dependent of $\Lambda$ and $\mu_{0}$.
    %$\theta$ can be written as the function of $\Lambda$ when $\mu_{0}$ is a given value.
    }
    \label{fig_theta}
\end{figure}

\section{Construction of neural-network ansatz based on the average fidelity} \label{ANN}
Recently a tentative scheme for the parameters is the usage of trigonometric functions as~\cite{kang_nonadiabatic_2022}
\begin{eqnarray}
    \mu & = & \mu_{0} + \Lambda \sin^{2} \bigg( \frac{\pi t}{L} \bigg) , \nonumber \\
    \eta & = & \eta_{0} + \pi \bigg[ 1 - \cos \bigg( \frac{\pi t}{L} \bigg) \bigg] ,
    \label{eq:mu_eta}
\end{eqnarray}
where $\Lambda$ is an auxiliary parameter depending on the concrete form of the desired gate.
% To implement the phase gate
% \begin{equation*}
%     \left[
%         \begin{array}{cc}
%             1 & 0 \\
%             0 & e^{i \phi}
%         \end{array}
%         \right],
% \end{equation*}
% we should make the off-diagonal elements of the evolution $U(T , 0)$ in Eq.~\eqref{eq:U} to become zeros.
% Thus, we set $\mu_{0} = 0$.
% In this case, $U(T , 0)$ has nothing to do with $\eta_{0}$.
% We choose $\eta_{0} = 0$ for simplicity.
% $\theta$ is set to be $- \phi / 2$ with ignoring the global phase.
% The corresponding $\Lambda$ can be yielded by Eq.~\eqref{eq:theta}.
% The last column in Tab.~\ref{table_fidelity} is the average fidelity of the corresponding phase gates.
To facilitate the subsequent discussion,
the parameter selection scheme of Eq.~\eqref{eq:mu_eta} is referred to as the trigonometric-function-based protocol.
We can numerically calculate the integral in Eq.~\eqref{eq:theta} as
\begin{eqnarray}
    \theta & = & \pi \bigg[ 1 - \sqrt{\frac{\pi}{2 \Lambda}} \Big( \cos(\mu_{0} + \Lambda) C(\sqrt{\frac{2 \Lambda}{\pi}}) \nonumber \\
    & & + \sin(\mu_{0} + \Lambda) S(\sqrt{\frac{2 \Lambda}{\pi}}) \Big) \bigg],
\end{eqnarray}
where
\begin{eqnarray*}
    S(x) & = & \int_{0}^{x} dt \sin(t^{2}),
    \quad C(x) = \int_{0}^{x} dt \cos(t^{2})
\end{eqnarray*}
are Fresnel integrals.
It is obvious that $\theta$ is only dependent on $\mu_{0}$ and $\Lambda$.
%When $\mu_{0}$ is known, it can yield all possible values of $\theta$ as $\Lambda$ varies.
%When we set $\mu_{0}$ to be $0$ , $\pi / 2$ and $\pi$,
%the changes of $\theta$ with $\lambda$ are shown in
We show the variation of $\theta$ with respect to $\Lambda$ for a few typical values of $\mu_{0}$ in Fig.~\ref{fig_theta}.
One observes that $\theta$ exhibits a decaying oscillation with respect to $\Lambda$ and
%It can be seen that $\theta$s change in similar ways with $\lambda$ when $\mu_{0}$ is given, only with modest differences in the amplitude and phase,
%and tend to %be a sine function
approaches $\pi$ when $\Lambda$ becomes sufficiently large. %as equilibrium position.
We find that $\theta$ cannot take %all values
the entire parameter range between 0 and $2 \pi$,
%, namely having a limited scope.
%Therefore,
which implies that the trigonometric-function-based protocol cannot implement arbitrary single-qubit gates.
Especially it is difficult to accurately obtain $\Lambda$ by solving complex nonlinear Eq.~\eqref{eq:theta}.
Therefore,
we improve the method of designing the variational parameters $\mu$ and $\eta$ by machine-learning-inspired optimization based on GRAPE.

%\red{Then, we use the neural network as an ansatz.}
Subsequently, we employ the neural network under unsupervised machine learning as an ansatz.
The neural network is composed of the input, hidden,
and output layers. %,  as is shown in Fig.~\ref{fig_ann}.
Two adjacent layers are connected by the weights, biases, and activation function.
We choose one hidden layer and $\tanh(x)$ as the activation function.
Because it is assumed that $\mu$ and $\eta$ have nothing to do with each other,
the neural network is not fully connected.
If the control parameters are not independent of each other,
the fully connected neural network will be adopted.
The final outputs are the specific function expressions
\begin{eqnarray*}
    \mu & = & \sum_{i = 1}^{N} W^{(2)}_{i} \tanh \Big(W^{(1)}_{i} \tau^{(1)} + B^{(1)}_{i} \Big) + B^{(2)}, \nonumber \\
    \eta & = & \sum_{i = 1}^{N} W^{(4)}_{i} \tanh \Big(W^{(3)}_{i} \tau^{(2)} + B^{(3)}_{i} \Big) + B^{(4)},
\end{eqnarray*}
where $N$ is the number of %the
neurons in the hidden layer.
Since the constructions of $\mu$ and $\eta$ are similar, %the same.
we take $\mu$ as an example.
$\tau^{(1)}$ is the input of the neural network.
The output of the neuron in the hidden layer is $\tanh \Big(W^{(1)}_{i} \tau^{(1)} + B^{(1)}_{i} \Big)$ with the weights $W^{(1)}_{i}$ and the biases $B^{(1)}_{i}$.
Similarly, the output of the neural network is the specific function expression of $\mu$ with the weights $W^{(2)}_{i}$ and the bias $B^{(2)}_{i}$.

\begin{figure}
    \includegraphics[width=\columnwidth]{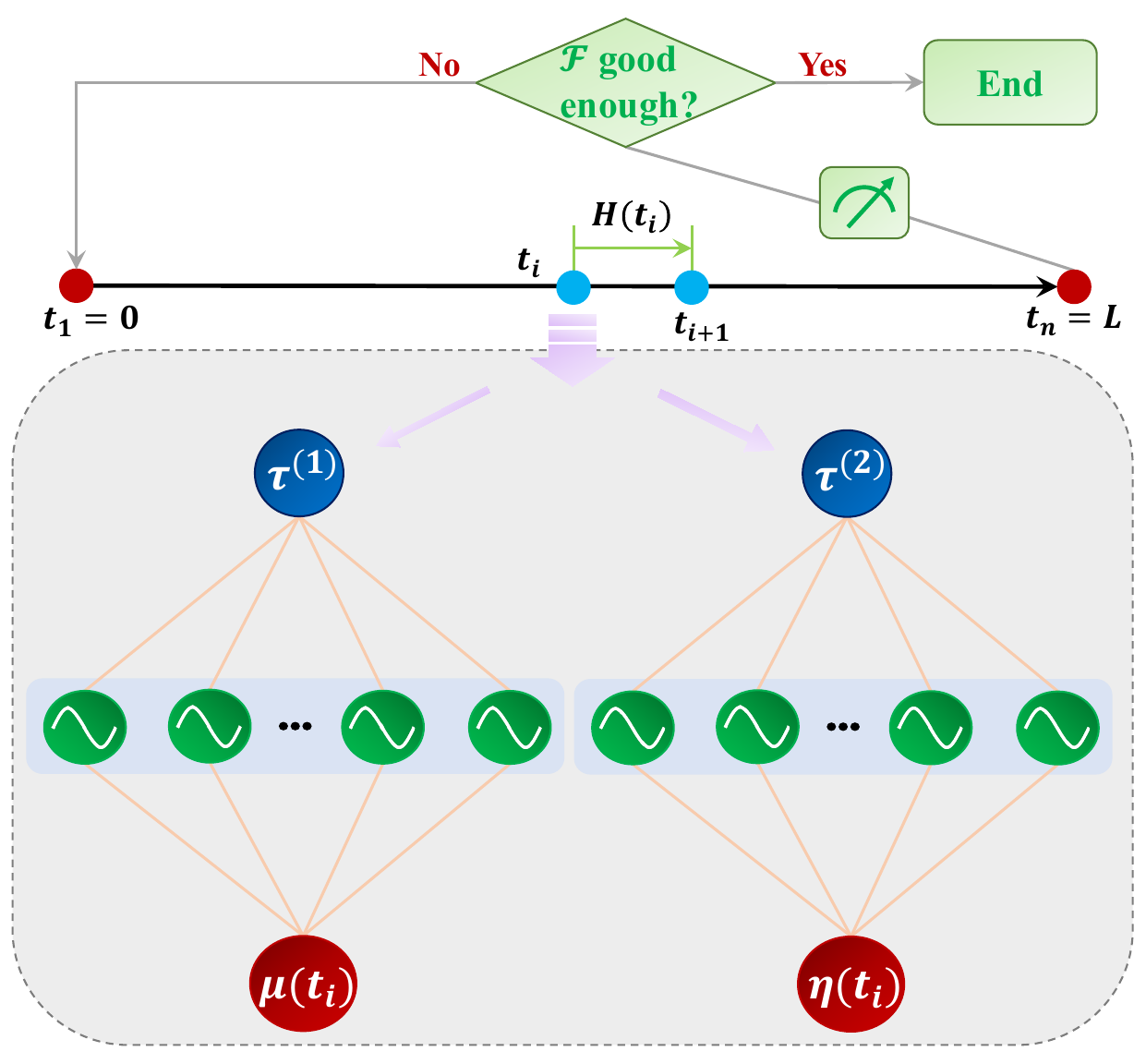}
    \caption{The workflow of the machine-learning-inspired optimization based on the average fidelity.
    For a temporal cycle between $t_{1} = 0$ and $t_{n} = L$ (red dots), which is divided into $n-2$ slices $\{ t_{i} \}_{i = 2}^{n-1}$ (blue dots), we should perform the variation and update all parameters at each time slice to ensure that the neural network captures the information at each moment effectively. % in order for the neural network to learn the information at each moment.
    Two adjacent dots are connected by the Schr\"{o}dinger equation, denoted as $U(t,0)$.
    We choose the neural network with one hidden layer. %(shaded area)
    $\tau^{(a)}$ ($a=1,2$) is the linear transformation of $t_{i}$ as the input and the output $\mu$ and $\eta$ are functions %with
    of $\tau^{(a)}$. % as independent variable.
    % $W$ and $B$ are weights and biases of each neuron, respectively.
    % Superscripts are used to distinguish %between
    % the hidden layer and the output layer.
    Each neuron in the hidden and output layers has a corresponding weight and bias.
    The trigonometric functions $\cos(\omega \tau^{(a)} + \phi)$ with different phases can be used as the inputs of the neurons in the hidden layer to ensure the output of the neural network is a periodic function,
    and $\omega$ is determined by the period of $\mu$ and $\eta$.
    Adjusting the bias of the output is useful to make $\mu$ and $\eta$ possess the fixed initial value.
    At the $t = L$ moment, the average fidelity is calculated according to the existing parameters to judge whether it is good enough to end the training.
    }
    \label{fig_ann}
\end{figure}

%\subsection
\textit{Feature enhancement.}
In parallel, we impose some restrictions on the variational parameters.
%initial values.
To meet the cycle evolution condition $| \phi_{\pm} (0) \rangle = | \phi_{\pm} (L) \rangle$,
the control parameters $\mu$ and $\eta$ should be periodic and $L$ is an integer multiple of the corresponding periods of $\mu$ and $\eta$.
Considering the period of $\mu$ is $T_{\mu}$ and
the period of $\eta$ is $T_{\eta}$,
%Therefore,
it is supposed that $T_{\eta} = m T_{\mu}$ with $m$ being any real number.
To be noticed, the periodicity of $\mu$ and $\eta$ is aimed at the real time $t$.
For simplicity,
we set $T_{\eta} = 2 T_{\mu} = L$.
%According to Eq.~\eqref{eq:U},
%$\mu_{0}$ and $\eta_{0}$ are determined for different gates.
%Thus, the control parameters have fixed initial values.
The initial values of the control parameters $\mu_{0}$ and $\eta_{0}$ can be determined by Eq.~\eqref{eq:U} for a target single-qubit quantum gate.

To summarize, the control parameters should meet three requirements below:
(1) $\mu$ and $\eta$ are periodic functions;
(2) $\mu$ and $\eta$ have initial value $\mu_{0}$ and $\eta_{0}$, respectively;
(3) $T_{\eta} = 2 T_{\mu} = L$.
The second condition can be satisfied easily. In particular,
$B^{(2)}$ and $B^{(4)}$ can be set depending on % \red{to let}
$\mu(0) = \mu_{0}$ and $\eta(0) = \eta_{0}$.
%We can let the input of $\mu$ to be $2 \pi t / L$ and the input of $\eta$ to be $\pi t / L$ to meet the third requirement.
To achieve the goal that $\mu$ and $\eta$ are periodic functions,
we ought to make periodic feature enhancement.

%For the purpose of the university, we take multi-layer neural network as an example.
Without loss of generality, we take the construction of a multi-layer neural network as an example. Considering the lemma that if $\iota(x)$ is a given smooth periodic function with period $L$ and $\Upsilon (\cdot)$ is a smooth function,
then $\Upsilon(\iota(x))$ is still a periodic function with period $L$~\cite{dong_method_2021}.
To proceed, we apply the sinusoidal functions
\begin{equation}
    \beta(x) = A \cos(\omega x + \phi) + c
\end{equation}
in the first hidden layer with $\omega =  2 \pi/L$.
We choose a nonlinear activation function for the sake of guaranteeing the periodicity of the output and generating higher-frequency terms to expand the model space
in training the neural network.
For other hidden layers,
the normal linear superposition of neurons in the former layer and nonlinear activation can be used.
%In practical use, we can design different structures of neural networks depending on a case.
%The more complex structure of the neural network behaves better at the expense of more time, for its further expanding the model space.
In this paper, we find that utilizing a small-scale neural-network ansatz with a single hidden layer is sufficient in optimizing the performance of target gates, which shows the superiority of our method.
However, it is worth noting that increasing the number of hidden units or incorporating additional hidden layers may yield improved behavior at the cost of increased computational time and more difficult physical realization.
In this respect, the final representations of $\mu$ and $\eta$ of the neural network with the sole hidden layer
are given by
\begin{eqnarray}
    \mu & = & \sum_{i = 1}^{N} W^{(2)}_{i} \tanh \Big[ W^{(1)}_{i} \cos \big(\omega^{(1)} \tau^{(1)} + \phi_{i}^{(1)} \big) + B^{(1)}_{i} \Big] \nonumber \\
    & & + B^{(2)},
    \end{eqnarray}
    \begin{eqnarray}
    \eta & = & \sum_{i = 1}^{N} W^{(4)}_{i} \tanh \Big[ W^{(3)}_{i} \cos \big(\omega^{(2)} \tau^{(2)} + \phi_{i}^{(2)} \big) + B^{(3)}_{i} \Big] \nonumber \\
    & & + B^{(4)}.
\end{eqnarray}
Here, $\tau^{(1)} = 2 \pi t / L$, $\tau^{(2)} = \pi t / L$,
and $\omega^{(1)} = \omega^{(2)} = 1$.
$\phi_{i}^{(1)}$ and $\phi_{i}^{(2)}$ are learnable parameters of the neural network,
which will effectively  expand the model space
and satisfy the periodic relationship between $\mu$ and $\eta$.

%\subsection
\textit{Backpropagation guided by the average fidelity.}
The average fidelity
is the benchmark to assess the performance of the quantum gates in the closed system and proves to be more effective than assessing the fidelity of specific states, especially in improving the performance of synthetic multi-qubit gates, given the uncertainties introduced by the preceding gate in the circuit.
We thus choose the average fidelity
as the objective function, %for improving the overall performance of the gates。 %instead of just optimizing the fidelity of \blue{specific states}.
which is give by~\cite{pedersen_fidelity_2007}
\begin{equation}
    F(t) = \frac{1}{\mathcal{D} (\mathcal{D} + 1)} \big[\Tr(M(t) M(t)^{\dagger}) + | \Tr(M(t)) |^{2}\big],
    \label{eq:f}
\end{equation}
%\red{which is more suitable to improve the performance of the synthetic multi-qubit gates due to the uncertainty of the previous gate in the quantum ciucuit.} \red{Here, }
where $\mathcal{D}$ is the dimension of the computational subspace,
$M(t) = \mathcal{P}_{c} U_{G}^{\dagger} U_{1}(t) \mathcal{P}_{c}$,
 $\mathcal{P}_{c}$ is the projective operators of the subspace,
and $U_{G}$ and $U_{1}(t)$ are the matrix representations of the ideal and actual gates, respectively.
The application of Eq.\eqref{eq:f} as the objective function instead of Eq.\eqref{eq:theta} offers two distinct advantages. First, Eq.\eqref{eq:f} takes into consideration the leakage to unwanted levels, making it a more realistic measure of the performance of the scheme compared to $\theta$, which is defined in the two-dimensional cat-state subspace.
%Secondly, it is widely acknowledged that neural networks tend to perform better with convex objective functions. Since the average fidelity is a convex function, it is a more suitable choice as an objective function for training the neural network compared to $\theta$, which has infinite ideal values that may potentially confuse the network~\cite{banchi_quantum_2016}.
%\red{Secondly, the average fidelity is non-convex functions~\cite{banchi_quantum_2016} given the parameter scale of the neural network, as well as $\theta$. The average fidelity has the established global maximum, which we can easily judge whether it is time to end the learning of the neural network, while $\theta$ has infinite ideal values that may potentially confuse the network. Therefore, it is a more suitable choice to treat the average fidelity as an objective function for training the neural network compared to $\theta$ and the fidelity of certain state.}
Secondly, while both the average fidelity and $\theta$ are non-convex functions~\cite{banchi_quantum_2016} due to the complex interactions among multiple parameters and the utilization of a nonlinear activation function, it is crucial to emphasize that there exists a clear global maximum for the average fidelity, which is unity. This allows for straightforward determination of when to finalize the neural network's learning process. On the other hand, $\theta$ encompasses infinite ideal values, which may potentially confuse the network.
Therefore, considering the aforementioned factors, employing the average fidelity as the objective function for neural network training is a more suitable choice compared to utilizing $\theta$ and the fidelity of specific states.
%\red{It is set that $\mathscr{F} = \bar{F}(L)$.
%The average fidelity takes the leakage to the unwanted levels into consideration,
%which is more realistic than $\theta$ to measure the performance of the scheme.
%Thus, we choose the average fidelity as the objective instead of $\theta$.
%Another reason for this is illustrated in Sec.~\ref{results}.}

The workflow of the machine-learning-inspired optimization is illustrated in Fig.~\ref{fig_ann}. In the neural-network ansatz, there are three layers with two input units, $N$ hidden units, and two output units.
%The specific connections between neurons are shown in Fig.~\ref{fig_ann}.
%
The final-state average fidelity $\mathcal{F} \equiv F(L)$ measured at the final moment depends crucially on the specific evolution details of each previous moment.
Considering the nonlinear relationship between the external single-photon drive in Eq.(\ref{eq:chi-epsilon}) and the control parameters,
it is challenging to directly derive the variation of $\mathcal{F}$ with respect to the neural-network parameters. Alternatively,  we use the greedy algorithm, in which the temporal period $L$ can be divided into $n$ discrete time slices during an evolution cycle of realizing a single-qubit gate. Optimizing the average fidelity at each time slice can lead to a substantial reduction in complexity, ultimately resulting in a higher overall average fidelity $\mathcal{F}$ for single-qubit gates.
It is obvious that %a link
the evolution between two contiguous moments is described by the Schr\"{o}dinger equation~\cite{khaneja_optimal_2005}.
%In this regard,
%we should  implement variation
%we have to apply variational methods at each discrete moment for acquiring higher final-state average fidelity $\mathcal{F}$ to realize single-qubit gates.
%\red{Due to the non-linear relationship between the control parameters and the control Hamiltonian, it is difficult to derive the variations if only the final-state fidelity $\mathcal{F}$ is chosen as the objective function.}
%It is obvious that %a link
%the evolution between two contiguous moments is described by Schr\"{o}dinger equation~\cite{khaneja_optimal_2005}.
%Then,
To this end, we calculate all the gradients of the average fidelity $F(t)$ with respect to parameters by the chain rule at each time slice.
In order to obtain the maximum of average fidelity $F(t)$,
we adopt the gradient ascent algorithm
%, in which we
to update all the parameters
\begin{eqnarray}
    W^{(a)} & \leftarrow & W^{(a)} + l_{W}^{(a)} \frac{\partial F(t)}{\partial W^{(a)}} , a = 1 , 2 , 3 , 4 ,\nonumber \\
    B^{(b)} & \leftarrow & B^{(b)} + l_{B}^{(b)} \frac{\partial F(t)}{\partial B^{(b)}} , b = 1 , 3, \nonumber \\
    \phi^{(c)} & \leftarrow & \phi^{(c)} + l_{\phi}^{(c)} \frac{\partial F(t)}{\partial \phi^{(c)}} , c = 1 , 2,
\end{eqnarray}
for the next variation.
Here, the learning rates $l_{W}^{(a)}$, $l_{B}^{(b)}$ and $l_{\phi}^{(c)}$ are the adjustable parameters,
depending on the impact of the corresponding parameters on the average fidelity $F(t)$.
We calculate the final-state average fidelity $\mathcal{F}$ corresponding to the current parameters
to judge whether the neural network %is
 has been well trained.
The process of the above operations is defined as one variational process. %processes $\mathscr{V}$.
After performing the variational processes $N_{\rm VP}$ times, we consider the training to be complete when $\mathcal{F}$ approaches 1 with high precision.
%To be noticed, only the shaded part is completed by the neural network, and the others are accomplished by the physical knowledge.
Note that the neural-network ansatz for the machine-learning-inspired method allows us to avoid solving Eq.~\eqref{eq:theta},
which has been assumed
%because it is considered
to be automatically met when the final-state average fidelity tends to unity.
%Therefore,
In this case, it is inevitable that we should verify whether Eq.~\eqref{eq:theta} is %true
valid or not according to the specific form of $\mu$ and $\eta$.

\begin{figure}
    \includegraphics[width=\columnwidth]{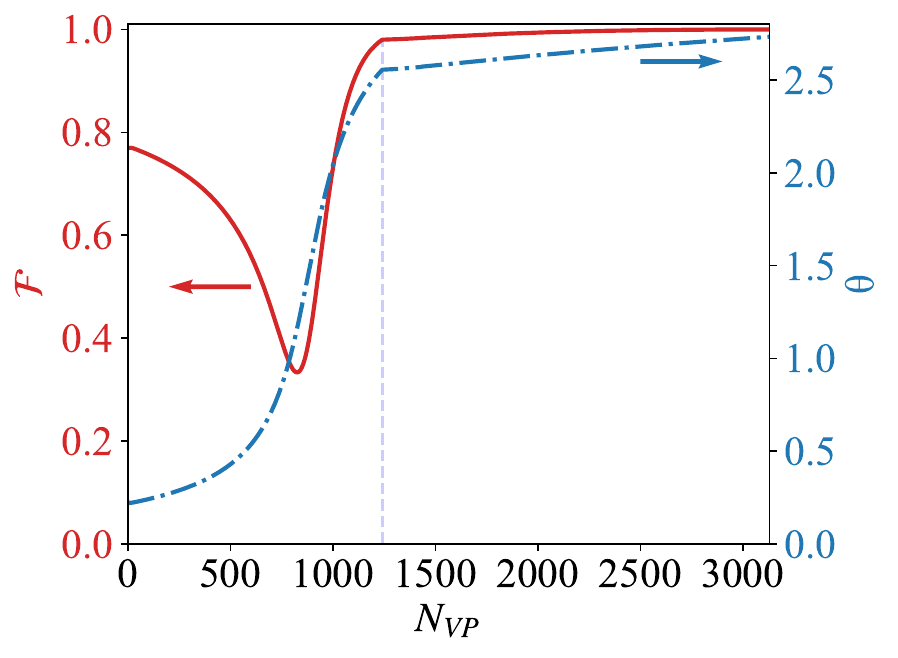}
    \caption{The training results of the T gate.
   The final-state average fidelity $\mathcal{F}$ (red solid line) and  $\theta$ (blue dash-dot line) change with the number of variational processes %$\mathscr{V}$.
   $N_{\rm VP}$.}
    %The blue line describes the changes of $\theta$ with variation.
    %The purple line represents where the learning rate changes.
    \label{fig_f}
\end{figure}

% \begin{figure}[htbp]
%     \centering

%     \subfigure[]{
%     \includegraphics[width=0.48\linewidth]{profig/5_7.pdf}
%     %\caption{fig1}
%     }%
%     % \quad
%     \subfigure[]{
%     \includegraphics[width=0.48\linewidth]{profig/5_8.pdf}
%     %\caption{fig2}
%     }%
%     \quad            %这个回车键很重要 \quad也可以
%     \subfigure[]{
%     \includegraphics[width=0.48\linewidth]{profig/5_9.pdf}
%     %\caption{fig2}
%     }%
%     % \quad
%     \subfigure[]{
%     \includegraphics[width=0.48\linewidth]{profig/5_10.pdf}
%     %\caption{fig2}
%     }%
%     \caption{The changes of $\mu(t)$ and $\eta(t)$ as the variation proceeds.
%     We take initial and final points and dots whose variational processes are 600, 1200, 1800, 2400 and 3000.
%     (a) and (b) described the variations of $\mu(t)$ and $\dot{\mu}(t)$ respectively.
%     (c) and (d) described the differences of $\eta(t)$ and $\dot{\eta}(t)$.
%     }
%     \label{fig_vary}
% \end{figure}

\section{Numerical results and discussion of single-qubit gates} \label{results}
The Gottesman-Knill theorem~\cite{PhysRevA.57.127}
tells us that a circuit using only Clifford gates and Pauli measurements~\cite{nielsen2002quantum} is insufficient for the universal quantum computation.
T gate%~\cite{nielsen2002quantum}
\begin{equation}
    \left[
        \begin{array}{cc}
            1 & 0 \\
            0 & e^{i \pi / 4}
        \end{array}
        \right]\\
\end{equation}
is the most natural and easiest single-qubit non-Clifford gate,
which supplements the set of Clifford gates to achieve universal quantum computation~\cite{boykin_new_2000, PhysRevA.86.022316}.
The implementation of the T gate in the trigonometric-function-based protocol is not perfect.
To realize the T gate,
we should make the off-diagonal elements of the evolution $U(T , 0)$ in Eq.~\eqref{eq:U} %to become zeros.
vanish.
Thus, we set $\mu_{0} = 0$.
In this case, $U(T , 0)$ has nothing to do with $\eta_{0}$.
We choose $\eta_{0} = 0$ for simplicity. For the
diagonal elements,
%we also should meet the conditions have
it is readily yielded that $2 k \pi - 2 \theta = \pi / 4$,
namely, $\theta = k \pi - \pi / 8$,
where $k$ is an arbitrary integer.

%\red{
%In the neural network, there are three layers with two input units, six hidden units and two output units.
%The specific connections between neurons are shown in Fig.~\ref{fig_ann}.
%}
In the neural network, there are six hidden units.
$\mu$ and $\eta$ use half of the hidden units, respectively, as shown in Fig.~\ref{fig_ann}.
We pre-train the neural network according to the trigonometric-function-based protocol to obtain the initial parameters.
Take a time series with $n = 1000$  data points evenly spaced between the time duration $t = 0$ and $L$.
Then, we set $l_{W}^{(2)} = l_{W}^{(4)} = 10^{-4}$, $ l_{W}^{(a)} = l_{B}^{(b)} = l_{\phi}^{(c)} = 10^{-5}$, $a = 1, 3$,  $b = 1, 3$, $c = 1, 2$ for the first 1240 iterations of variational processes,
and $l_{W}^{(2)} = l_{W}^{(4)} = 10^{-5}$, $l_{W}^{(a)} = l_{B}^{(b)} = l_{\phi}^{(c)} = 10^{-6}$, $a = 1, 3$,  $b = 1, 3$, $c = 1, 2$ for the last 1890 iterations of variational processes.
The learning rates of $W^{(2)}$ and $W^{(4)}$ are ten times more than those of other parameters,
because the impact of $W^{(2)}$ and $W^{(4)}$ on the average fidelity $F(t)$ is much larger than other parameters.
In the NGQC+,
the amplitude of coherent states is $| \alpha | = 0.5$,
the Kerr nonlinearity is $K = 2 \pi \times 12.5 \rm{MHz}$,
the energy gap is $E_{gap} = 4 K \alpha^{2} = 78.5 \rm{MHz}$~\cite{grimm_stabilization_2020},
and the total interaction time is $T = 1 \mu s$.
For simplicity, we refer to our scheme as the machine-learning-inspired protocol.

\begin{figure}
    \includegraphics[width=\columnwidth]{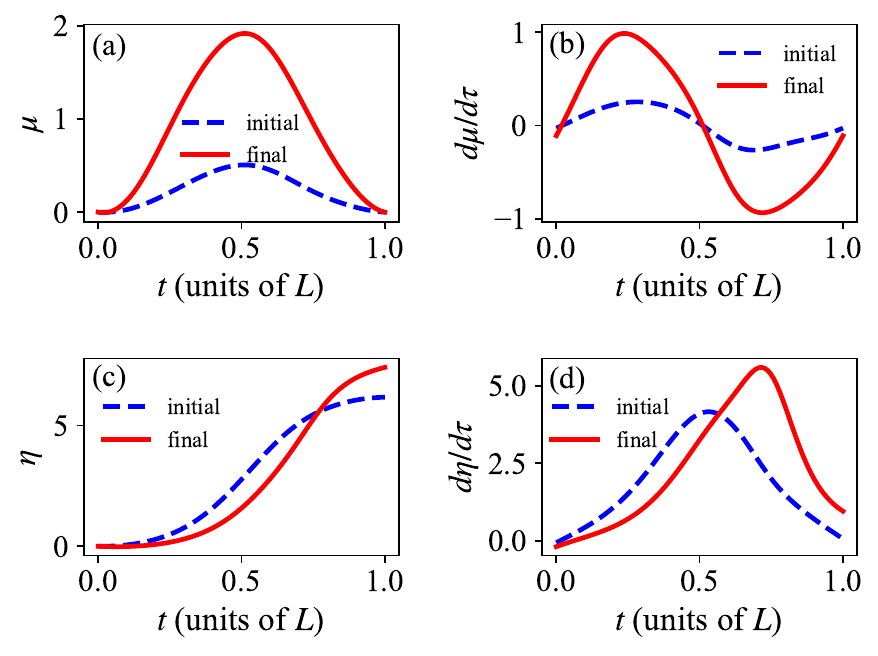}
    \caption{
    %The changes of $\mu$ and $\eta$ as the variation processes. %proceeds.
    %We plot the
    The comparison between the initial and the final forms of (a) $\mu$ and (c) $\eta$.  The initial forms take trigonometric functions,
    %which are the outcomes of pre-training the neural network,
    and  the final forms are the outputs of the neural network.
    The first derivatives with respect to the inputs of the neural network (b) $d \mu / d\tau^{(1)}$ and
    (d) $d \eta / d\tau^{(2)}$ are also compared.}
     %respectively.
    \label{fig_vary}
\end{figure}

The training results of the T gate  through the code in Ref.\cite{mao_neural_2023} are shown in Fig.~\ref{fig_f}.
The final-state average fidelity is 0.9999.
According to Eq.~\eqref{eq:theta},
we get $\theta = 2.7332$,
and the error from the theoretical value ($\theta = 7 \pi / 8$) is $0.0156$.
It can be seen that the trends of the final-state average fidelity and $\theta$ are different.
%At the beginning of the variational processes,
During the initial several hundred iterations of the variational processes,
the final-state average fidelity is moderately high,
with the corresponding $\theta$ value already nearing another ideal value of $- \pi / 8$.
%while the corresponding $\theta$ is close to another ideal value $- \pi / 8$.
%\red{It is acknowledged that the neural network prefers concave-convex functions as the objective functions.
%The average fidelity is a convex function,
%while $\theta$ has infinite ideal values, which may confuse the neural network.}
%So the benchmark,
The final-state average fidelity will approach unity and simultaneously the actual value of $\theta$ converges to the ideal value
as the variational iterations progress.
%after undergoing the all the variational iterations.
More importantly, Fig.~\ref{fig_f} illustrates the advantage of employing unsupervised learning in this paper. Unsupervised learning does not rely on a predefined set of labeled data for guidance, unlike supervised learning.
Instead, it relies solely on gradients for learning, without the notion of right or wrong during the initial stages. As a result, it may initially go in the wrong direction. However, with proper initialization, it can eventually find the correct direction for learning.
One can infer that the average fidelity
is superior to the imposed constraint of $\theta$ in Eq.~\eqref{eq:theta}.
%useful to judge whether Eq.~\eqref{eq:theta} is satisfied. It is reasonable to omit conditional constraints [Eq.~\eqref{eq:theta}] in the process of network design.
Thus, it is wise to choose the average fidelity instead of $\theta$ as the objective function.

%because
%In fact, the neural network is much easier to solve the non-integral problem [citation].
% and the average fidelity has the uniquely determined ideal value.
%\red{It also display the characteristic of non-supervised learning.
%Unlike the supervised learning, it is only guided by the gradients without the judgment of right and wrong,
%and it may go the wrong direction at the beginning.
%However, it can find the right direction in the final if it is well initialized.}

As such,
we %choose
show the comparison between the initial and final points
%to study how
of $\mu(t)$ and $\eta(t)$ %vary as the variational processes go on,
%as shown
in Fig.~\ref{fig_vary} (a) and %Fig.~\ref{fig_vary}
(c).
The initial forms take trigonometric functions,
which are the outcomes of pre-training the neural network.
The final forms are obtained by the outputs of the neural network.
%The red lines are trigonometric functions,
%which are the outcomes of pre-training the neural network.
%The blue lines are the output of the neural network.
%It can be seen
One finds that %the amplitude and symmetry vary.
the final form has a clear deviation in the amplitude and the structure symmetry from the initial form after the entire training of the neural network.
%We find that the maximum points will have a certain deviation from the original positions,
%and they
The final forms are no longer simple trigonometric functions, which can be clearly revealed
%The above conclusions can be more obvious through
by the derivatives of $\mu(t)$ with respect to $\tau^{(1)}$ and $\eta(t)$ with respect to $\tau^{(2)}$ shown in Fig.~\ref{fig_vary} (b) and (d).
%The derivatives are no longer symmetrical,
%and the zero points of the derivatives shift.
%These are all evidences that they are no longer simple trigonometric functions.
%Thus, t
The introduction of the neural network can broaden the model space,
%from the linear space to the nonlinear space.
in which the control parameters can %obtain
take more extensive and feasible trial forms.

\begin{figure}
    \includegraphics[width=\columnwidth]{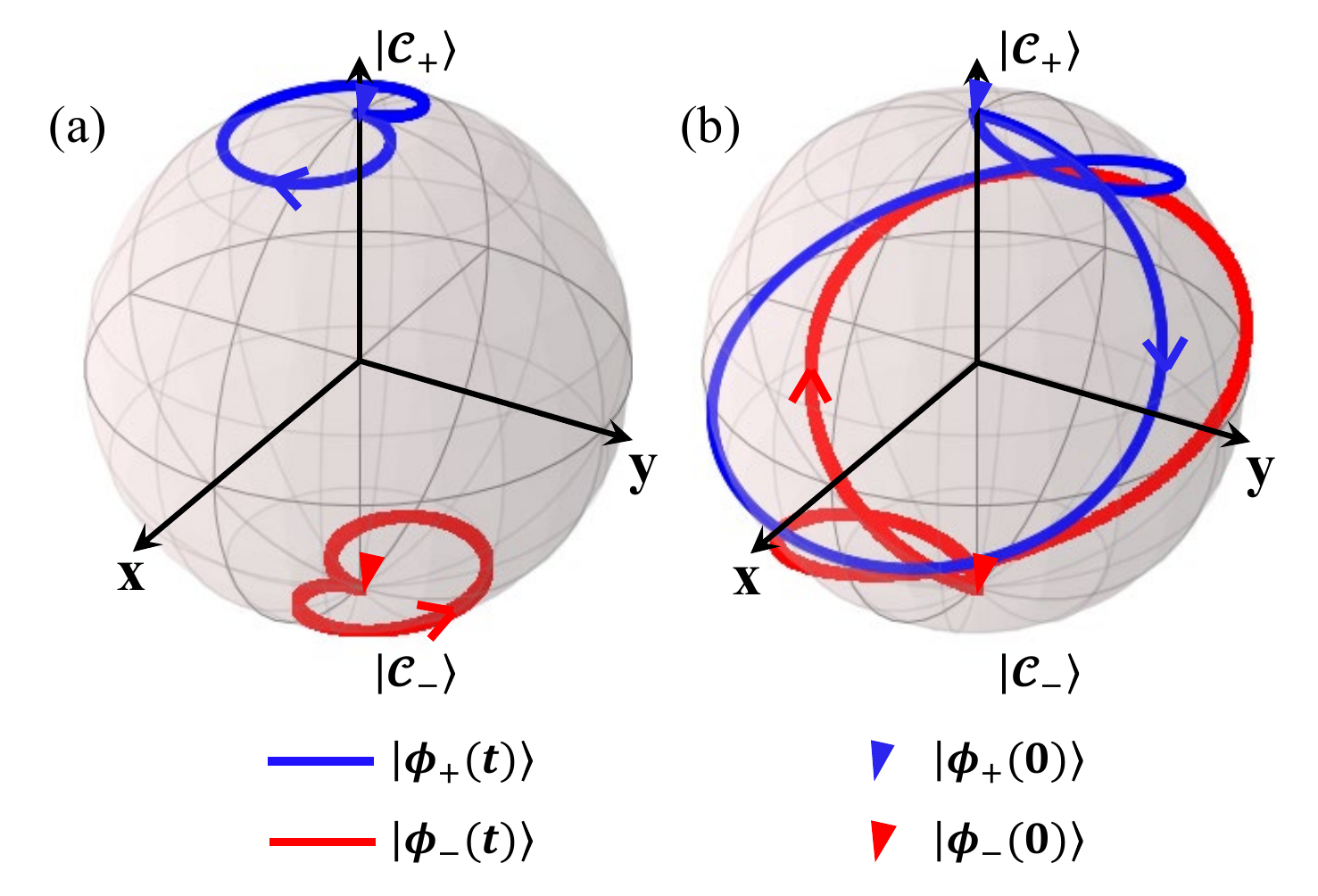}
    \caption{Map of the changes of $\mu$ and $\eta$ into the changes of the evolution path of the eigenstates $| \phi_{\pm}(t) \rangle$ on the Bloch sphere.
    The blue line is the evolution paths of $| \phi_{+}(t) \rangle$.
    The red line is the evolution path of $| \phi_{-}(t) \rangle$.
    $| \mathcal{C}_{\pm} \rangle$ are the initial states of the evolution of $| \phi_{\pm}(t) \rangle$.
    (a) The evolution path corresponds to the initial $\mu$ and $\eta$.
    (b) The evolution path corresponds to the final $\mu$ and $\eta$.
    }
    \label{fig_bloch}
\end{figure}

To get more insights into the behaviors
%According to the
of $\mu(t)$ and $\eta(t)$ at the initial and final points of the training,
we plot the trajectories of the eigenstates $|\phi_{\pm}(t)\rangle$ on the Bloch sphere in Fig.~\ref{fig_bloch}:
\begin{equation}
    \vec{r}_{\pm}(t) = \sum_{k = x , y , z} \mathrm{Tr} \big[ | \phi_{\pm} \rangle \langle \phi_{\pm} | \sigma_{k} \big] \vec{e}_{k},
\end{equation}
where $\vec{e}_{k}$ is the unit vector along the $k$ axis.
The differences between the initial and final $\mu(t)$ and $\eta(t)$ are magnified on the Bloch sphere.
It can be seen that the evolution path varies a lot during the entire training.
Thus,
the neural-network ansatz %does a good job
shows unique advantages
in quantum optimal control,
which can obtain a more complex ansatz for possible control parameters.
%and possesses the powerful ability of the parameter representations.

\begin{figure}
    \includegraphics[width=\columnwidth]{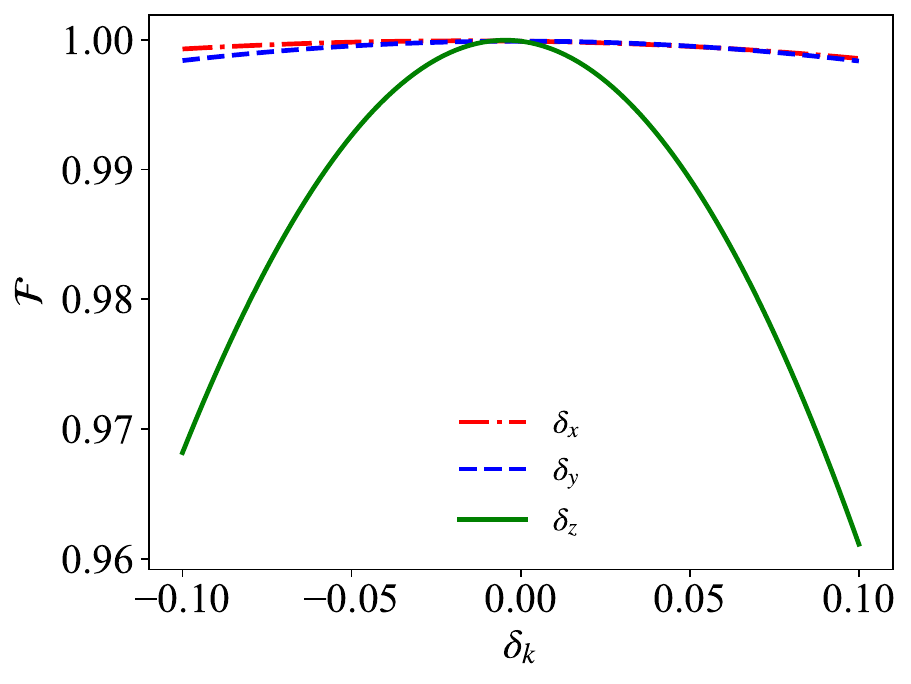}
    \caption{The variation of final-state average fidelity $\mathcal{F}$ of the T gate with respect to the systematic error coefficient $\delta_{k}$, $k = x, y , z$.}
     %under the systematic noise effect.
    \label{fig_systematic}
\end{figure}

\textit{Noise robustness.}
Next,
we evaluate the performance of our scheme under different noisy circumstances.
First,
we consider the systematic noise effect,
%Due to the
such as instrument defects and imperfection operations.
%The existence of systematic error induces the average value of measured data deviates significantly from the ideal %value. one.
Systematic errors can cause the average value of measured data to deviate significantly from the ideal value.
%The characteristic of systematic error is that the measurement result changes according to a certain rule, with repeatability and unidirectionality.
%Therefore,
The influence of systematic errors may be present in
the parameters of the control Hamiltonian that can be written as $\Omega_{k}^{e} = (1 + \delta_{k}) \Omega_{k}$, $k = x , y , z$,
where $\delta_{k}$ is the error coefficient.
We plot the final-state average fidelity $\mathcal{F}$ of the T gate with respect to the error coefficient $\delta_{k}$ in Fig.~\ref{fig_systematic}.
We can find that when $\delta_{x} \in [-0.1 , 0.1]$ $(\delta_{y} \in [-0.1 , 0.1])$,
the final-state average fidelity $\mathcal{F}$ remains higher than 0.9986 (0.9984),
    while we can only obtain $\mathcal{F} \ge 0.9611$ when $\delta_{z} \in [-0.1 , 0.1]$.
It is obvious that the noise in the $z$ axis direction will cause more catastrophic decline in the final-state average fidelity than that in $x$ and $y$ axes. %axis.
This effect can be understood because according to Eq.~\eqref{eq:dp}
the fluctuation in $\Omega_{z}$ will cause persistent adverse effects on the dynamic phase.
The unidirectional offset of $\Omega_{z}$ will make the dynamic phase not vanish after a cycle,
and then spoils the conditions of NGQC+.

\begin{figure}[t!]
    \includegraphics[width=\columnwidth]{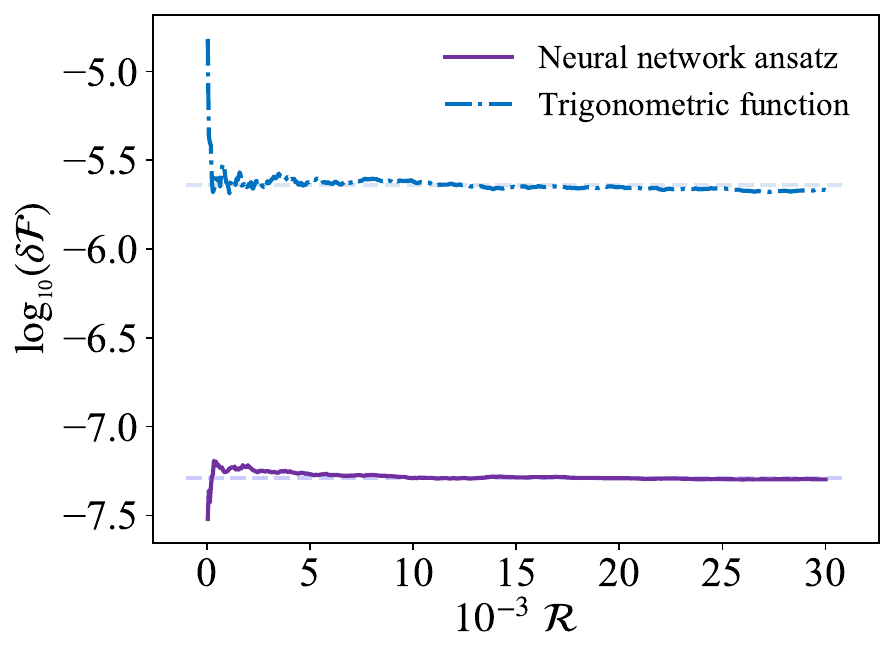}
    \caption{The logarithms of the deviations $\delta \mathcal{F}$ of the mean values of final-state average fidelities of the T gate from the ideal value with respect to simulation times $\mathcal{R}$ under the random noise effect with $\rm{SNR} = 10$
    in the machine-learning-inspired (purple solid line) and trigonometric-function based (blue dash-dot line) protocols.
    The dotted lines represent the convergence values of the two protocols.
    Here, $\mathcal{R} = 50p , (p = 1 , 2 , 3 , \cdots)$.
    }
    \label{fig_random}
\end{figure}

We also consider the random noise effect,
in which the amplitude, waveform, and phase are random at any time.
Each random noise is still subject to certain statistical distribution.
If the amplitude distribution of a noise follows Gaussian distribution and its power spectral density is uniformly distributed,
this noise is called additive white Gaussian noise (AWGN).
AWGN is one of the typical random noise models.
Therefore, we take AWGN as an example to analyze the robustness of our method to random processes
and compare the robustness of the machine-learning-inspired and trigonometric-function based protocols.
We add the AWGN to control parameters as
\begin{equation}
    \Omega_{k}^{q}(t) = \Omega_{k}(t) + \mathcal{A}_{G} [\Omega_{k}(t) , \rm SNR],
\end{equation}
where $q$ represents each random generator of AWGN,
$\mathcal{A}_{G}[\Omega_{k}(t) , \rm SNR]$ is a function that generates AWGN for the original signal $\Omega_{k}(t)$
with signal-to-noise ratio $\rm SNR = 10 \log_{10} (P_{signal} / P_{noise})$,
and $\rm{P_{signal}}$ and $\rm{P_{noise}}$ are the power of signal and noise, respectively.
Due to the random generation of AWGN,
we perform a large amount of numerical simulations to estimate the random noise effect.
The logarithms of the deviations $\delta \mathcal{F}$ of the mean values
of final-state average fidelities
of the T gate from the ideal value
 of $50p$ $(p = 1 , 2 , 3 , \cdots)$ %times
iterations of numerical simulations are plotted in Fig.~\ref{fig_random} with $\rm{SNR} = 10$.
When $p$ tends to infinity,
the simulation consequence is pretty close to the actual impact of the random noise.
The ideal value of the final-state average fidelity in the machine-learning-inspired protocol is 0.9999,
while that in the trigonometric-function-based protocol is 0.8894.
%One observes that our scheme behaves better under the random noise.
It is observed that our scheme performs significantly better in the presence of random noise.
Compared to the trigonometric-function-based protocol,
the mean value of the final-state average fidelities under random noise in the machine-learning-inspired scheme exhibits %more scarce
fewer fluctuations with respect to $p$ and approaches $1 - 7.84 \times 10^{-4}$,
with a smaller deviation from the ideal value as $p$ becomes sufficiently large.
It is thus acknowledged that our scheme is robust against the random noise,
and enhancing the performance of the gate can improve the robustness to a certain degree.

\begin{table}[b!]
    \caption{Parameters and corresponding final-state average fidelities for the implementation of single-qubit gates.
    The rightmost three columns are to verify whether Eq.~\eqref{eq:theta} is satisfied.
    We calculate $\theta_{\rm actual}$ with the output of the neural network,
    and the relative error between $\theta_{\rm actual}$ and $\theta_{\rm ideal}$.}
    \label{table_gate} %table的label要插在caption之后
    \centering
    \begin{tabular}{ccccccc}
        \hline
        \hline
        Gate & $\mu_{0}$ & $\eta_{0}$ & Fidelity & $\theta_{\rm actual}$ & $\theta_{\rm ideal}$ & Error \\
        \hline
        $\rm{T}$ & 0 & 0 & 0.9999 & 2.7332 & $7 \pi / 8$ & 0.0157 \\
        $\rm{X}$ & $3 \pi / 2$ & $\pi / 2$ & 0.9999 & 1.5459 & $ \pi / 2$ & 0.0249 \\
        $\rm{H}$ & $\pi / 4$ & $\pi / 2$ & 0.9997 & 1.5738 & $\pi / 2$ & 0.0030 \\
        $\rm{T^{\dagger}}$ & 0 & 0 & 0.9999 & 0.3569 & $\pi / 8$ & 0.0338 \\
        $\rm{R_{x}(\pi / 4)}$ & $\pi / 2$ & $- \pi /2$ & 0.9992 & 3.6019 & $9 \pi / 8$ & 0.0676 \\
        \hline
        \hline
    \end{tabular}
\end{table}

As the system cannot be completely isolated from the environment,
the inevitable interactions between the system and the environment will also lead to the decoherence.
We mainly consider two dissipation factors,
such as a single-photon loss and dephasing~\cite{harrington_engineered_2022}.
The evolution of the system can be described by the Lindblad master equation~\cite{harrington_engineered_2022, mirrahimi_dynamically_2014}:
\begin{eqnarray}
    \dot{\rho}(t) & = & - i [H_{\rm cat} + H_{c}(t) , \rho(t)] \nonumber \\
    & & + \Gamma \mathcal{L} [a] \rho(t) + \Gamma_{\phi} \mathcal{L} [a^{\dagger} a] \rho(t) .
\end{eqnarray}
Here, $\Gamma$ and $\Gamma_{\phi}$ are the dissipation coefficients of a single-photon loss and dephasing,
respectively,
and the Lindblad superoperator $\mathcal{L}$ acting on arbitrary operator $o$ produces $\mathcal{L}[o] \rho(t) = o \rho(t) o^{\dagger} - o^{\dagger} o \rho(t) / 2 - \rho(t) o^{\dagger} o / 2$.
In the presence of decoherence,
the evolution is no more unitary.
We can no longer use Eq.~\eqref{eq:f} to measure the performance of the quantum gates.
Therefore,
we take the evolution with initial state $| \mathcal{C}_{+} \rangle$ as an example and evaluate the fidelity of the T gate as
\begin{equation}
F_{T} =\left\langle \mathcal{C}_{+} \Big| U_{T}^{\dagger}
\rho(T) U_{T} \Big| \mathcal{C}_{+} \right\rangle.
\end{equation}
In our numerical simulation,
we set $\Gamma = \Gamma_{\phi} = 0.05 \rm{MHz}$,
and we can obtain the fidelity of the T gate as $F_{T} = 0.9803$.
This means the leakage to unwanted levels outside the subspace is still very small,
and our scheme is insensitive to decoherence.

The implementations of the $\rm{NOT}$ gate ($\rm{X}$ gate), Hadamard gate ($\rm{H}$ gate), $\rm T^{\dagger}$ gate, and $\rm{R_{x}(\pi / 4)}$ gate are listed in Tab.~\ref{table_gate}.
Here, $\rm{R}_{x}(\phi) = \exp (- \frac{i}{2} \phi \sigma_{x} )$
%\begin{equation}
%R_{x}(\phi) = \exp \Big(- \frac{i}{2} \phi \sigma_{x} \Big)$,
%\end{equation}
is a rotation gate around the $x$-axis~\cite{nakanishi2021quantum}.
It can be seen that the machine-learning-inspired protocol %scheme can do a great job
excels
for almost all kinds of single-qubit gates.
Especially,
our scheme %performs wonderfully
shows superiority in phase gates,
whose average fidelities can reach 0.9999,
much higher than those in the trigonometric-function-based protocol.
The performance of the X gate in the two protocols is equally remarkable.
%The performances of X gate in the two protocol are equally wonderful.
Through the neural network,
we can implement the rotation gates which are unrealizable in the trigonometric-function-based protocol.
For the $\rm{H}$ gate,
the obtained results are not very accurate, and more sophisticated neural networks are awaited.
Furthermore,
we realize the modified controlled-NOT $(\rm{CNOT})$ gate with the final-state average fidelity 0.9996.
Here, $U_{\rm{CNOT}} = | \mathcal{C}_{+} \rangle \langle \mathcal{C}_{+} | \otimes \mathbb{I} + | \mathcal{C}_{-} \rangle \langle \mathcal{C}_{-} | \otimes (- i \sigma_{x})$ and
$\mathbb{I}$ is the unit matrix acting on the cat-state subspace.
%How to excute in principle
The execution of a two-qubit controlled gate is shown in Appendix~\ref{two-qubit}.
It is clear that for each gate, the higher the final-state average fidelity is,
the smaller the error is.

%it can obtain a result slightly worse than the original one.

To conclude,
through the introduction of the neural network,
we can %break
lift the restrictions imposed on $\theta$ to a certain extent.
%In theory,
We can realize arbitrary $\theta$ by adjusting the initial parameters and the structure of the neural network.

\section{The realization of Toffoli gate} \label{toffoli}

\begin{figure}[t!]
    \includegraphics[width=\columnwidth]{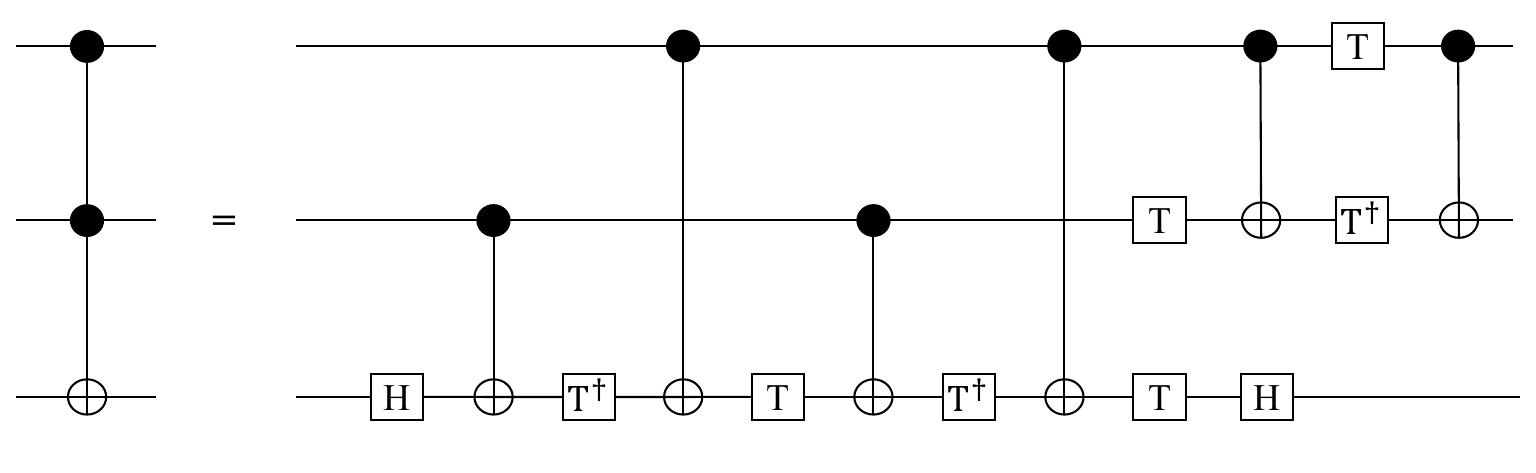}
    \caption{The Toffoli gate is composed of the CNOT gate, H gate, T gate ($\pi / 8$), and $\rm T^{\dagger}$ ($- \pi / 8$) gate.
    }
    \label{fig_toffoli}
\end{figure}

In the trigonometric-function-based protocol,
%it is scarcely possible to execute the synthetic or
it is scarcely possible to execute the single-shot multi-qubit gates, and the final-state average fidelity of synthetic multi-qubit gates by combining high-fidelity single- and two-qubit gates will be rather low.
The Toffoli gate~\cite{nielsen2002quantum}
\begin{equation}
    \left[
        \begin{array}{cccccccc}
            1 & 0 & 0 & 0 & 0 & 0 & 0 & 0 \\
            0 & 1 & 0 & 0 & 0 & 0 & 0 & 0 \\
            0 & 0 & -1 & 0 & 0 & 0 & 0 & 0 \\
            0 & 0 & 0 & -1 & 0 & 0 & 0 & 0 \\
            0 & 0 & 0 & 0 & 1 & 0 & 0 & 0 \\
            0 & 0 & 0 & 0 & 0 & 1 & 0 & 0 \\
            0 & 0 & 0 & 0 & 0 & 0 & 0 & -1 \\
            0 & 0 & 0 & 0 & 0 & 0 & -1 & 0
        \end{array}
        \right] \nonumber
\end{equation}
is composed of the CNOT gate, H gate, T gate, and $\rm T^{\dagger}$ gate,
as is shown in Fig.~\ref{fig_toffoli}.
The final-state average fidelity of Toffoli gate is 0.5169,
when all gates in Fig.~\ref{fig_toffoli} are realized in the trigonometric-function-based protocol,
and the main limitation is due to the bad performance of the T gate.
In the machine-learning-inspired protocol, we can realize higher-fidelity multi-qubit gates in the cascaded mode.
The finely modified Toffoli gate can be
%with taking the specific realization of single- and two-qubit gates into consideration.
%Exploit
synthesized by H gate and CNOT gate implemented in the trigonometric-function-based protocol
and T gate and $\rm T^{\dagger}$ gate %discussed in Sec.~\ref{results},
shown in Tab.\ref{table_gate}, and the final-state average fidelity of such a three-qubit entangling gate %shown in Fig.~\ref{fig_toffoli}.
increases to 0.9976,  with improved performance of the $\rm{T}$ and $\rm{T^{\dagger}}$ gates. However, we find that,
%The perfection of $\rm{T}$ and $\rm{T^{\dagger}}$ gates contributes a lot to the implementation of the Toffoli gate.
%while that of Toffoli gate in original scheme is 0.5169.
%it can be seen that
although the final-state average fidelities of $\rm{T}$ and $\rm{T^{\dagger}}$ gates are up to 0.9999,
it is still challenging to synthesize a high-fidelity multi-qubit gate, which is seriously hindered by the lowest-fidelity quantum gates.
When all gates in Fig.~\ref{fig_toffoli} are realized in the machine-learning-inspired protocol,
the final-state average fidelity can further increase to 0.9981.
%We think there is still room for improvement.
%still fails to meet the standard of realizing the synthesized multi-qubit gates.
%We think realizing the synthetic multi-qubit gates requires that the average fidelities of most of the components should reach $99.99\%$.
Thus, our scheme can provide a feasible routine to realize multi-qubit gates in bosonic systems.

\section{Conclusion and outlook} \label{conclusion}

In this paper,
we present a machine-learning-inspired method of optimizing the performance of the imperfect gates with cat-state NGQC+ via reverse engineering.
By utilizing periodic feature enhancement and corresponding biases,
we can obtain a periodic function as an output of a neural-network ansatz with fixed initial values.
The machine-learning-inspired protocol allows us to not have to solve the difficult nonlinear equation [Eq.~\eqref{eq:theta}],
%Besides, it is
which can be automatically satisfied when the final-state average fidelity tends to be 1. Through analyzing the variational forms of the control parameters and comparing with the simple trigonometric functions,
we prove that the neural network can greatly expand the model space and realize a more complex ansatz for possible control parameters.
We find the final-state average fidelities of the phase gates and $\rm{NOT}$ gate can reach 0.9999,
and those of the Hadamard gate and CNOT gate can be up to 0.9996.
%In the hope of improving
In order to improve the performance of the Hadamard gate,
we can expand the scale of the neural network by increasing the number of hidden units and %the number of
hidden layers.
%in order to improve the fitting ability.
We can also adjust the periodic relationship between $\mu$ and $\eta$ and the initial parameters in the hope of obtaining better results.
%It is insightful to make $\theta$ as the objective function to modify the parameters of the neural network.
An alternative approach in the neural network is to use multi-objective optimization.
%The neural network with multi-objective optimization is also an alternative.
Meanwhile,  we show that we developed an approach for implementing high-fidelity rotation gates that are challenging to realize using trigonometric function-based protocol. Our scheme demonstrates robustness against various types of decoherence effects.
%we can realize the rotation gates that are scarcely possible in the trigonometric-function-based protocol.
%Moreover, we find that when the average fidelities of single- and two-qubit gates reach a certain level, the average fidelities of the synthetic gate may not be too low.
Additionally, we observe that once the average fidelities of single- and two-qubit gates surpass a certain threshold, the average fidelities of the synthetic gate may not be significantly compromised.
%the increase of the average fidelities may not lead to the increase of the average fidelity of the synthetic gate.
%Combining the trigonometric function and our scheme,
Combining high-fidelity single- and two-qubit gates,
we can implement the Toffoli gate with high fidelity,
which can not be simply realized in trigonometric-function-based protocol. In order to further improve the performance of the synthetic gate,
we %will
can use the average fidelity of the synthetic gate to guide the variational learning of the neural network,
instead of only optimizing the single and two-qubit gates, and the improved scheme is left for a future study.
We thus provide an alternative method of designing the control parameters.
The machine-learning-inspired scheme % smooths the path
paves the way for the optimization of continuous and periodic parameters in the quantum control,
and can be generalized to more intricate neural networks featuring a substantial number of optimizable parameters,
targeting increasingly complex quantum systems~\cite{PhysRevA.102.022617, liu_optimized_2021, li_invariant-based_2021, Wang_2023}.
%\red{in which the scale of the neural-network ansatz can be flexibly modified case by case.
%The optimization works more effectively in our scheme under the circumstances that the objective function is physical at each time slice}.
%In our scheme,

%Moreover,
%we can also introduce the decoherence effects into our scheme.
%We need to use the master equation as a bridge connecting two contiguous times,
%and find a new benchmark to guide the variational learning of the neural network.

\begin{acknowledgments}
The authors appreciate very insightful discussions with Yimin Wang, Ming Xue and Meng-Jiao Lyu.
This work is supported by College students' innovation and
entrepreneurship training program projects of Nanjing University of
Aeronautics and Astronautics under Grant 202210287094Z.
W.-L. Y. kindly acknowledges support by the National Natural Science
Foundation of China (NSFC) under Grant No. 12174194 and a startup fund
of Nanjing University of Aeronautics and Astronautics under Grant No.
1008-YAH20006.
\mbox{A.M.O. kindly} acknowledges Narodowe Centrum Nauki
(NCN, Poland) Project No. 2021/43/B/ST3/02166 and is grateful
for support via the Alexander von Humboldt \mbox{Foundation}
\mbox{Fellowship} \mbox{(Humboldt-Forschungspreis).}
\end{acknowledgments}

\appendix
\section{The realization of two-qubit controlled gate} \label{two-qubit}
The Hamiltonian of two cavity modes driven by two Kerr-nonlinear resonators can be described as
\begin{equation}
    H_{\rm cat,2} = \sum_{n = 1, 2} \big(- K a_{n}^{\dagger 2} a_{n}^{2} + \epsilon_{2} (e^{2 i \xi} a_{n}^{\dagger 2} + e^{- 2 i \xi} a_{n}^{2}) \big).
\end{equation}
Here, $a_{n}$ $(a^{\dagger}_{n})$ is the annihilation (creation) operator of the $n$th mode.% $a_{n}$.
The product states of two-mode coherent states %of two modes
$\{| \alpha \rangle_{1} \otimes | \alpha \rangle_{2}\}$ ,
with $\alpha = \pm \sqrt{\epsilon_{2} / K} \exp(i \xi)$,
are four-fold degenerate eigenstates of $H_{\rm cat2}$.
$\{ | \mathcal{C}_{\pm} \rangle_{1} \otimes | \mathcal{C}_{\pm} \rangle_{2} \}$ can span the four-dimensional subspace $\mathcal{S}_{2}$ to implement the two-qubit gates.
The control Hamiltonian~\cite{mirrahimi_dynamically_2014, PhysRevX.9.041053, PhysRevLett.122.080502} is given by
\begin{eqnarray}
    H_{c2}(t) & = & \chi_{12}(t) a_{1}^{\dagger} a_{1} a_{2}^{\dagger} a_{2} + a_{1}^{\dagger} a_{1} \Big[\lambda^{\ast}(t) a_{2} + \lambda(t) a_{2}^{\dagger}\Big] \nonumber \\
    & + & \epsilon^{\ast}(t) a_{2} + \epsilon(t) a_{2}^{\dagger} + \sum_{n = 1 , 2} \chi_{n}(t) a_{n}^{\dagger} a_{n}.
\end{eqnarray}
Here, $\chi_{12}(t)$ is the cross-Kerr parameter,
$\lambda(t)$ is the longitudinal interaction strength between modes 1 and 2,
$\epsilon(t)$ is the strength of the extra driving of mode 2,
and $\chi_{n}$ ($n$=1,2) is the detuning of the $n$th mode.
Similarly, it is assumed that the parameters of $H_{c2}$ should be much smaller than the energy gaps between cat states and other eigenstates of $H_{\rm cat,2}$.
To realize the two-qubit controlled gate $U_{2} (T , 0) = | \mathcal{C}_{+} \rangle_{1} \langle \mathcal{C}_{+} | \otimes \mathbb{I}_{2} + | \mathcal{C}_{-} \rangle_{1} \langle \mathcal{C}_{-} | \otimes U_{s}(T , 0)$,
the parameters of $H_{c2}$ are set as follows:
\begin{eqnarray}
    \chi_{12}(t) & = & \frac{- 2 \Omega_{z} \mathcal{N}_{+}^{2} \mathcal{N}_{-}^{2}}{\mid \alpha \mid^{4} (\mathcal{N}_{+}^{2} - \mathcal{N}_{-}^{2})^{2}} , \nonumber \\
    \chi_{1}(t) & = & - \frac{\mid \alpha \mid^{2} (\mathcal{N}_{+}^{2} + \mathcal{N}_{-}^{2})}{2 \mathcal{N}_{+} \mathcal{N}_{-}} \chi_{12}(t) , \nonumber \\
    \chi_{2}(t) & = & - \mid \alpha \mid^{2} \frac{\mathcal{N}_{-}}{\mathcal{N}_{+}} \chi_{12}(t), \nonumber \\
    \rm{Re}[\lambda(t)] & = & \frac{(\mathcal{N}_{+} \mathcal{N}_{-})^{\frac{3}{2}}}{4 (\mathcal{N}_{+}^{2} - \mathcal{N}_{-}^{2}) \mid \alpha \mid^{3}} (\Omega_{x} \cos \xi - \Omega_{y} e^{2 \mid \alpha \mid^{2}} \sin \xi), \nonumber \\
    \rm{Im}[\lambda(t)] & = & \frac{(\mathcal{N}_{+} \mathcal{N}_{-})^{\frac{3}{2}}}{4 (\mathcal{N}_{+}^{2} - \mathcal{N}_{-}^{2}) \mid \alpha \mid^{3}} (\Omega_{x} \sin \xi + \Omega_{y} e^{2 \mid \alpha \mid^{2}} \cos \xi), \nonumber \\
    \rm{Re}[\epsilon(t)] & = & - \mid \alpha \mid^{2} \frac{\mathcal{N}_{-}}{\mathcal{N}_{+}} \rm{Re}[\lambda(t)] , \nonumber \\
    \rm{Im}[\epsilon(t)] & = & - \mid \alpha \mid^{2} \frac{\mathcal{N}_{-}}{\mathcal{N}_{+}} \rm{Im}[\lambda(t)] ,
\end{eqnarray}
which are slightly different from the parameters chosen in Ref.~\cite{kang_nonadiabatic_2022}.
The optimization of two-qubit controlled gates in the neural networks is similar to that of single-qubit gates.

\normalem %% Do not delete
\bibliographystyle{apsrev4-2}
%\bibliography{prorefs1}


\end{document}